# Comparing evaluation of quantitative imaging methods using reference standards vs. regression-without-truth-based technique

Yan Liu[a], Abhinav K. Jha[a,b,*]

[a]Department of Biomedical Engineering, Washington University in St. Louis, St. Louis, MO, USA;
[b]Mallinckrodt Institute of Radiology, Washington University in St. Louis, St. Louis, MO, USA

## ABSTRACT

Clinical translation of quantitative imaging (QI) methods requires objective evaluation of these methods on reliably measuring the underlying true quantitative values. Ideally, such evaluation would be performed using ground truth or gold standards. However, in clinical practice, ground truth is generally unavailable, and obtaining gold standards for large patient cohorts is often impractical or infeasible. In these scenarios, evaluation using well-accepted or commonly used methods for measuring the quantitative values that may have bias and/or measurement error, referred to as reference standards, is a common alternative. However, the error in reference standards might lead to inaccurate results in evaluating the QI methods. To address the challenge of evaluation without a gold standard, a class of regression-without-truth (RWT)-based techniques has also been proposed. These techniques compare QI methods without requiring ground truth and have demonstrated promise across multiple applications. This raises an important question: how does evaluation using reference standards vs. RWT-based techniques compare? Our work, which investigates this question using controlled numerical experiments, showed that when the measurement error of the reference standard exceeded a certain level, evaluation with a RWT-based technique outperformed evaluation with reference standards. This study provides insights for selecting appropriate evaluation strategies when gold standards are unavailable and motivates further investigation in clinically realistic settings.

## 1. INTRODUCTION

Quantitative imaging (QI), the process of extracting numerical or statistical features from medical images to support diagnosis, treatment planning and therapy monitoring, has shown increasing promise in clinical practice[1,2]. Examples include single-photon emission computed tomography (SPECT)-based dosimetry in radionuclide therapy[3], positron emission tomography (PET)-derived metabolic tumor volume (MTV) for predicting response to cancer therapy[4], and computed tomography-derived tumor heterogeneity for monitoring treatment response[5]. The important clinical value of these applications has led to the development of numerous QI methods[6–8]. For clinical translation, objective evaluation of these methods with respect to their ability to reliably measure the underlying quantitative values is essential.

When ground truth is available, such as in virtual imaging trials[9] and physical phantom studies[10], measured quantitative values can be directly compared with ground truth for objective evaluation of QI methods. However, in patient studies, ground truth is generally unavailable. In such cases, evaluation using a gold standard may be considered, where a gold standard is defined as the most reliable available method for measuring the quantitative value of interest and is presumed to be correct. In practice, however, obtaining gold-standard data for large patient cohorts is often difficult or infeasible.

In the absence of gold standards, evaluation with reference standards is commonly used as an alternative, where reference standards are defined as well-accepted or commonly used methods for measuring the underlying quantitative value but have associated bias and/or measurement error[11]. Such error in the reference standards could potentially introduce error in the comparison of QI methods.

Another set of techniques that have been recently developed to perform evaluation in the absence of gold standards are a class of regression-without-truth (RWT)-based techniques[12–15]. These techniques are based on the premise that the measured values are result of specific image formation, reconstruction and quantification process applied to the true value.

*a.jha@wustl.edu

Consequently, the measured and true values should be statistically related. Specifically, RWT-based techniques assume a linear relationship between the true and measured values, characterized by a slope, a bias and a zero-mean Gaussian distributed noise term. With the assumption that the true values are sampled from a parametric distribution, a maximum-likelihood approach can be derived to estimate the linear-relationship parameters without access to the true values. These techniques then use the ratio of the estimated noise standard deviation to slope, referred to as noise-to-slope ratio, to compare different QI methods in terms of precision. These techniques showed promise in applications across different modalities. Examples include comparing different PET segmentation methods for quantifying metabolic tumor volume[16], different image reconstruction methods in SPECT for measuring mean regional activity uptake[14,17] and different segmentation methods in cardiac cine MRI for measuring cardiac ejection fraction[18].

Given potential impact of measurement error in reference standards and the demonstrated promise of RWT-based techniques, an important question arises: How do these two kinds of evaluation strategies compare in their ability to accurately rank QI methods. In this manuscript, we examined this question through numerical experiments. Our studies quantitatively compare RWT-based techniques with evaluation with reference standards. These investigations were conducted for varying degrees of error in the reference standard.

## 2. METHODS

### 2.1 Regression-without-truth-based technique

Consider $K$ QI methods are used to measure certain quantitative values from the images of a patient population consisting of $P$ patients. For the $p^{th}$ patient, the true quantitative value is denoted as $a_p$, the measured value yielded by the $k^{th}$ QI method is denoted as $\hat{a}_{p,k}$. For the $k^{th}$ method, RWT-based techniques assume a linear relationship between the measured values and the true values. This linear relationship holds due to imaging physics and quantification process in multiple cases. For example, imaging systems such as PET and SPECT can be described using linear operators. Several image reconstruction algorithms are linear, and even certain iterative reconstruction approaches can be reasonably approximated as linear[19]. Finally, the process of quantification from reconstructed images is also linear for several features such as regional activity uptake. In fact, a test of linear relationship between measured and true quantitative values is often an important step in evaluating quantitative imaging biomarkers, as it allows a change in the true value be reflected proportionally in the measured value an average[20]. We characterize this linear relationship by a slope $u_k$, a bias $v_k$ and a zero-mean Gaussian noise term.

In this study, we consider a specific RWT-based technique, named NGSE-Corr (no-gold-standard evaluation technique that models correlated noise). This technique models the correlated noise across different QI methods[15]. Specifically, the correlated noise of different methods is modeled using a multi-variate Gaussian distribution, denoted as $\mathcal{N}(0, \Sigma)$. The diagonal elements of $\Sigma$, i.e., $\{\sigma_k^2\}$, denote the variance of the noise of each method. The off-diagonal elements of $\Sigma$, i.e., $\{\sigma_{k,k'}\}$, denote the covariance of the noise between methods $k$ and $k'$. Thus, for the $p^{th}$ patient, the relationship between true and measured values can be written as

$$\begin{bmatrix} \hat{a}_{p,1} \\ \hat{a}_{p,2} \\ \vdots \\ \hat{a}_{p,K} \end{bmatrix} = \begin{bmatrix} u_1 & v_1 \\ u_2 & v_2 \\ \vdots & \vdots \\ u_K & v_K \end{bmatrix} \begin{bmatrix} a_p \\ 1 \end{bmatrix} + \mathcal{N}(0, \Sigma). \tag{1}$$

Denote the measurements yielded by $K$ QI methods for the $p^{th}$ patient, $\{\hat{a}_{p,k}, k = 1, \ldots, K\}$, by $\widehat{\boldsymbol{A}}_p$, the matrix containing $\{u_k\}$ and $\{v_k\}$ by $\boldsymbol{\Theta}$, denote $[a_p, 1]^T$ by $\hat{A}_p$ Based on Eq. 1, we can write the probability distribution of $\widehat{\boldsymbol{A}}_p$ given the true value $a_p$ as:

$$\text{pr}(\widehat{\boldsymbol{A}}_p | \boldsymbol{\Theta}, \boldsymbol{\Sigma}, a_p) = \mathcal{N}(\boldsymbol{\Theta}\hat{A}_p, \boldsymbol{\Sigma}), \tag{2}$$

where $\text{pr}(x)$ denotes the probability of a random variable $x$.

Next, NGSE-Corr assumes that the true value $a_p$ is sampled from a distribution parameterized by a vector $\boldsymbol{\Omega}$, denoted as $\text{pr}(a_p | \boldsymbol{\Omega})$. In that case, we can marginalize over the true values $a_p$, yielding

$$\text{pr}(\widehat{\boldsymbol{A}}_p | \boldsymbol{\Theta}, \boldsymbol{\Sigma}, \boldsymbol{\Omega}) = \int \text{pr}(\widehat{\boldsymbol{A}}_p | a_p, \boldsymbol{\Theta}, \boldsymbol{\Sigma}) \text{pr}(a_p | \boldsymbol{\Omega}) da_p. \tag{3}$$

By further assuming the true values of *P* patients are independent of each other, we can obtain the log-likelihood function for all the measurements. Denoting this log-likelihood by $\Lambda(\Theta, \Sigma, \Omega | \{\widehat{A}_p\})$, we obtain:

$$\Lambda(\Theta, \Sigma, \Omega | \{\widehat{A}_p\}) = \sum_{p=1}^{P} \ln \int \mathrm{pr}(\widehat{A}_p | a_p, \Theta, \Sigma) \mathrm{pr}(a_p | \Omega) da_p. \tag{4}$$

NGSE-Corr uses a maximum likelihood approach to estimate parameters $\{\Theta, \Sigma, \Omega\}$. After obtaining the estimates, we note that the bias and the slope can be recalibrated for the measurements. This recalibration process results in the standard deviation of the normally distributed noise term to be divided by the slope term. Since this noise term cannot be subtracted, it quantitatively measures the performance of the QI methods on the basis of precisely quantifying the measured values. Thus, the ratio of estimated noise standard deviation to estimated slope, termed as noise-to-slope ratio (NSR), is computed as a figure of merit to rank the QI methods based on precision. Lower NSR value indicates more precise QI method.

## 2.2 Evaluation with reference standards

As mentioned earlier, the reference standards can have errors. In our study, we considered these errors to be stochastic, i.e. no bias in the reference standard was assumed. Following Obuchowski et al[11], we model the reference standard measurements $\tilde{a}_p$ as unbiased, with measurement error characterized by a zero-mean Gaussian noise with standard deviation $\sigma_\omega$. Specifically, the reference standard measurement for the $p^{th}$ patient, denoted by $\tilde{a}_p$, was given by

$$\tilde{a}_p = a_p + \mathcal{N}(0, \sigma_w^2). \tag{5}$$

In patient studies, the true values $\{a_p\}$ are not observed, and only the reference standards $\{\tilde{a}_p\}$ are available. In this case, the reference standards are used as surrogate for the true values. The linear relationship parameters are estimated by performing least-squares regression between the QI method measurements and the reference standards. The resulting estimated noise-to-slope ratio is then used to rank the QI methods based on precision.

## 2.3 Numerical study

A numerical study was conducted to investigate the performance of the two evaluation strategies. We first generated a set of true quantitative values sampled from a known parametric distribution. Based on these true values, measurements were simulated for three hypothetical QI methods. The simulated measurements were then input into the NGSE-Corr technique to estimate the linear relationship parameters by maximizing the log-likelihood function in Eq. (4). Using the estimated parameters, the noise-to-slope ratio (NSR) was computed to rank the three hypothetical QI methods.

For evaluation using reference standards, reference measurements were generated by adding zero-mean Gaussian noise to the true values in Eq. (5). These reference measurements were treated as surrogates for the true values. Linear regression was then performed between the QI method measurements and the reference measurements to estimate the corresponding linear relationship parameters. The estimated noise-to-slope ratio was subsequently used to rank the QI methods based on precision.

To conduct the analysis in a clinically relevant context, we used linear relationship parameters derived from a realistic simulation where three quantitative SPECT methods were used to measure the mean regional activity uptake in patients with prostate cancer treated with $^{223}$Ra. The number of patient samples was varied for both NGSE-Corr and evaluation using reference standards. In addition, for evaluation using reference standards, the noise standard deviation was varied to investigate the impact of different levels of measurement noise in the reference standards on ranking the methods. For each number of patients, 50 noise realizations were simulated to compute the accuracy of correctly ranking the methods and identifying the most precise method.

## 3. RESULTS

Fig. 1(A) compares the evaluation with reference standards and RWT-based technique, specifically NGSE-Corr, in correctly ranking the QI methods. For evaluation with reference standards, different levels of measurement noise were introduced. When the noise standard deviation of the measurement error in reference standard was relatively small, evaluation with reference standards outperformed NGSE-Corr. However, we observe that as the noise standard deviation in references standard increased, NGSE-Corr achieved superior ranking performance comparing to evaluation with reference standards. Similar results can be obtained in correctly identifying the most precise method as shown in Fig. 1(B).

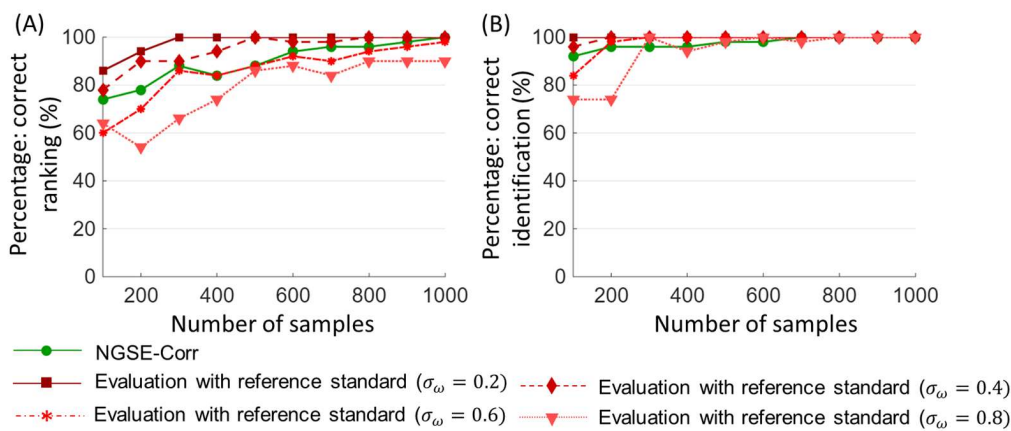

Figure 1: A comparison of NGSE-Corr (green lines) and evaluation with reference standard (red lines) of (A) correctly ranking all the QI methods and (B) correctly identifying the most precise method.

## 4. CONCLUSION AND DISCUSSION

In this study, we conducted a numerical study to compare the performance of evaluation using reference standards and a RWT-based technique for ranking QI methods in the absence of ground truth. The results of our studies showed that as measurement noise in the reference standards increases, the performance of evaluation with reference standards in correctly ranking the methods can become inferior to the considered RWT-based technique. This study provides preliminary evidence suggesting that evaluation using reference standards may yield lower ranking accuracy than RWT-based techniques under certain conditions. These findings also have some implications for evaluating QI methods in clinical scenarios where ground truth is unavailable.

One such potential context is evaluating PET segmentation methods on the task of quantifying metabolic tumor volume (MTV). In clinical settings, physician delineations are often used as reference standards to compare image segmentation methods. However, these delineations are subject to intra-observer and inter-observer variability, which may introduce error into the evaluation process. In such scenario, RWT-based techniques may be potentially used to evaluate different segmentation methods. Thus, our results motivate comparing physician-delineation-based evaluation vs. RWT-based techniques for comparing PET segmentation methods.

Another relevant context is the evaluation of PET denoising methods for quantitative tasks such as quantifying MTV or total lesion glycolysis. To improve patient comfort, reduced PET acquisition time has been explored. However, shorter scan durations result in low-count PET images with increased noise. To address this issue, various PET denoising methods have been developed[21]. In this case, measurements from normal count PET images are commonly used as reference standards. However, measurements from normal count images are not ground truth and they still suffer from noise, which could potentially affect the evaluation results. In this case, RWT-based technique could potentially provide an alternative approach for comparing denoising methods. Thus, the results of this study motivate a comparison between normal-count-image-based evaluation vs. RWT-based techniques for evaluating PET denoising methods.

A limitation of this study is that the analysis was conducted using numerical simulations. The simulated data were generated under controlled conditions and satisfied the assumptions of the underlying mathematical models. In clinical practice, however, these assumptions may not hold. Reference standards may exhibit more complex error structures and may not be adequately represented as ground truth with additive Gaussian noise. Therefore, an important next step is to compare reference-standard-based evaluation vs. RWT-based technique for evaluating QI methods in clinically realistic scenarios.

In conclusion, findings from our numerical studies to compare evaluation of QI methods using reference standards versus regression-without-truth-based techniques provide preliminary insight into selecting evaluation strategies when ground truth is unavailable and motivate further investigation under clinically realistic conditions.

## ACKNOWLEDGEMENTS

This work was supported by the National Institute of Biomedical Imaging and Bioengineering of the National Institute of Health under grants R01-EB031051, R01-EB031962 and NSF CAREER Award 2239707.